\documentclass[fleqn,usenatbib]{mnras}

\usepackage{newtxtext,newtxmath}

\usepackage[T1]{fontenc}

\DeclareRobustCommand{\VAN}[3]{#2}
\let\VANthebibliography\thebibliography
\def\thebibliography{\DeclareRobustCommand{\VAN}[3]{##3}\VANthebibliography}

\usepackage{graphicx}	
\usepackage{amsmath}	
\usepackage{subcaption}
\usepackage{xcolor}

\newcommand{\gtsimeq}{\raisebox{-0.6ex}{$\,\stackrel{\raisebox{-.2ex}{$\textstyle >$}}{\sim}\,$}}

\title[The mm Fundamental Plane of XRBs]{Stellar-mass black holes on the millimetre fundamental plane of black hole accretion}

\author[J.\ S.\ Elford et al.]{\parbox{\textwidth}{
Jacob S.\ Elford,$^{1,2}$\thanks{E-mail: elfordj@cardiff.ac.uk}
Ilaria Ruffa,$^{1,3}$
Timothy A.\ Davis,$^{1}$
Martin Bureau,$^{4}$
Rob Fender,$^{4}$
Jindra Gensior,$^{5}$
Thomas Williams,$^{4}$
Hengyue Zhang $^{4}$
} 
\newauthor
\\
$^{1}$Cardiff Hub for Astrophysics Research \&\ Technology, School of Physics \&\ Astronomy, Cardiff University, Queens Buildings, The Parade, Cardiff, CF24 3AA, UK\\
$^{2}$Instituto de Estudios Astrof\'{i}sicos, Facultad de Ingenier\'{i}a y Ciencias, Universidad Diego Portales, Av.\ Ej\'{e}rcito Libertador 441, Santiago, Chile\\
$^{3}$INAF, Arcetri Astrophysical Observatory, Largo Enrico Fermi 5, I-50125 Florence, Italy\\
$^{4}$Sub-department of Astrophysics, Department of Physics, University of Oxford, Keble Road, Oxford, OX1 3RH, UK\\
$^{5}$Institute for Astronomy, University of Edinburgh, Royal Observatory, Blackford Hill, Edinburgh EH9 3HJ, UK\\
}

\date{Accepted XXX. Received YYY; in original form ZZZ}

\pubyear{2015}

\begin{document}
\label{firstpage}
\pagerange{\pageref{firstpage}--\pageref{lastpage}}
\maketitle

\begin{abstract}
Recent work revealed the existence of a galaxy "millimetre fundamental plane of black hole accretion", a tight correlation between nuclear $1$~mm luminosity, intrinsic $2$ -- $10$~keV X-ray luminosity and supermassive black hole mass, originally discovered for nearby low- and high-luminosity active galactic nuclei.
Here we use mm and X-ray data of $5$ X-ray binaries (XRBs) to demonstrate that these stellar-mass black holes also lie on the mm fundamental plane, as they do at radio wavelengths. One source for which we have multi-epoch observations shows evidence of deviations from the plane after a state change, suggesting that the plane only applies to XRBs in the hard state, as is true again at radio wavelengths. We show that both advection-dominated accretion flows and compact jet models predict the existence of the plane across the entire range of black hole masses, although these models vary in their ability to accurately predict the XRB black hole masses.
\end{abstract}



\section{Introduction}
Supermassive black holes (SMBHs; with masses $>10^6$~M$_\odot$) are now known to lie at the centre of every galaxy with a total stellar mass $M_\star>10^9$~M$_\odot$ and are believed to co-evolve with their hosts (see e.g.\ \citealp{KormendyHo2013} for a review). The many details of such co-evolution are however still poorly understood \citep[see e.g.][]{Donofrio21}. To better understand the SMBH-host galaxy connection it is key to understand how the accretion physics of these objects and whether the same physics holds across all mass regimes including down to stellar mass BHs.

The so-called “fundamental plane of BH accretion” (heareafter FP) is an empirical correlation between the SMBH masses ($M_{\rm BH}$), $5$-GHz radio ($L_{\rm 5GHz}$) and $2$ -- $10$-keV X-ray ($L_{\rm X,2-10}$) luminosities of galaxies, which was initially reported by \citet{Merloni03} and \citet{Falcke04}. This correlation is supposed to arise due to the physics in the regions surrounding the SMBHs, and thus provides a probe of several different processes. However, this radio FP has been notoriously hard to interpret, with both radio jets and shocks touted as potential driving mechanisms. Furthermore, the scatter around this correlation (intrinsic scatter $\approx0.9$~dex for all galaxies and $\approx0.6$~dex for low-luminosity active galactic nuclei, LLAGN; see e.g.\ \citealp{Merloni03}, \citealp{Gultekin19} and references therein) is much larger than that around the $M_{\rm BH}$ -- $\sigma_\star$ (stellar velocity dispersion) relation ($\approx0.3$~dex), the gold standard of SMBH -- galaxy correlations.

Atacama Large Millimeter/sub-millimeter Array (ALMA) observations have recently revealed an analogous ``millimetre fundamental plane of BH accretion'' \citep{Ruffa24}, where the integrated/large-scale radio continuum luminosity is replaced by the nuclear (only) $230$-GHz millimetre (hereafter mm) continuum luminosity. The mm fundamental plane of BH accretion (mmFP) was discovered for a sample of $48$ sources primarily drawn from the mm-Wave Interferometric Survey of Dark Object Masses (WISDOM) project. These $48$ sources are all at redshifts $z\lesssim0.03$, span a large range of AGN bolometric luminosities ($L_{\rm bol}=10^{41}$ -- $10^{46}$~${\rm erg~s^{-1}}$) and mostly (but not exclusively) have very low rates of accretion onto their central SMBHs ($\dot{M}\lesssim10^{-3}$~$\dot{M}_{\rm Edd}$, where $\dot{M}_{\rm Edd}$ is the Eddington accretion rate; \citealp{Elford24}). Crucially, the intrinsic scatter about this plane ($\approx0.4$~dex) is comparable to that of the $M_{\rm BH}-\sigma_{\star}$ relation \citep[e.g.][]{Vandenbosch16} and can be smaller than that of its radio counterpart (depending on the sample used to fit the plane; e.g.\ \citealp{Merloni03,Falcke04,Gultekin09,Plotkin12,Gultekin2019}). When restricting the analysis of the mmFP to only those sources with the most accurate (redshift-independent) distances, an even tighter correlation is obtained \citep[see][]{Ruffa24}, with an intrinsic scatter ($\approx0.11$~dex) comparable to that of some of the tightest scaling relations in astronomy (such as the baryonic Tully-Fisher relation; e.g.\ \citealp{Lelli16,Lelli19}). Surprisingly, the observed correlation holds for both LLAGN and sources from the {\it Swift}-BAT AGN Spectroscopic Survey (BASS), the latter comprising some of the brightest and most powerful nearby AGN (with median $L_{\rm bol}\approx10^{44}$~${\rm erg~s^{-1}}$ and $\dot{M}\approx0.01$ -- $0.1$~$\dot{M}_{\rm Edd}$; \citealp{Koss17}), hinting at a greater AGN homogeneity than previously thought. 

Models of advection-dominated accretion flows (ADAFs) and compact radio jets seem to be able to explain the mmFP, while classical torus models cannot \citep[see][]{Ruffa24}. This new empirical correlation thus provides both a proxy (i.e.\ an indirect estimate) of SMBH mass and crucial new insights into SMBH accretion physics. In a typical low-luminosity (i.e.\ low-accretion rate) AGN, a classic accretion disc should be either absent or truncated at some inner radius (the transition usually happening beyond a few tens of Schwarzschild radii), and should be replaced by some sort of ADAFs \citep{Nara95,Ho08}. Compact (i.e. unresolved) jets have been argued to dominate the whole SEDs of LLAGN \citep[e.g.][]{Fernandez-Ontiveros2023} and have spectral properties similar to those of ADAFs at radio and mm wavelengths. There is evidence that these mechanism may be key play a role in other types of accreting object with \citet{Ichimaru77} suggested that the observed spectra of hard-state X-ray binaries such as Cygnus X-1 are likely dominated by emission from ADAF-like accretion processes. Additionally, it has been found that the radio luminosities of compact jets from XRBs show a strong, non-linear correlation with their X-ray luminosities (e.g.\ \citealp{Corbel03} and \citealp{Gallo03} and references therein). It is important to test whether stellar-mass BHs in accreting systems (i.e.\ X-ray binaries, XRBs) also follow the mmFP (as they do the radio FP) which could hint at an even greater universality of accretion mechanisms.

In this paper, we present an analysis of the mmFP of stellar-mass black holes, by using both archival and new data of 5 XRBs. Our main aim is to establish whether the newly-discovered mmFP holds also for accreting stellar-mass black holes, and if it does whether ADAF models can also explain the correlation of this type of sources, as they seem to do for AGN of all accretion rates.

\section{mm and X-ray observations of stellar-mass black holes}
\label{sec:methods-fitting}

Only a few tens of XRBs have confirmed stellar-mass black holes. Most of these are truly transients, with a range of several orders of magnitude in X-ray luminosity from quiescence to outburst. Very few among the confirmed accreting stellar systems remain detected even in quiescence, varying by only a factor of $\sim10$ over decades. Strikingly, such sources follow the same XRB radio -- X-ray correlation, that becomes the radio FP with the addition of a (BH) mass term \citep[e.g.][]{Merloni03,Gultekin09}. Crucially, and unlike SMBHs, an individual XRB can be tracked as it moves back and forth across the plane, showing the fundamental plane can hold for individual objects and samples of objects over different mass scales. These correlations are exhibited during the so-called ``hard state'', when a strong (accretion flow) coronal X-ray emitting component causes the correlation with the radio jet. Accreting stellar-mass black holes in this phase are considered scaled-down equivalents of LLAGN. At high luminosities, XRBs (and likely also AGN) may enter a ``soft'' accretion state, when the jet is suppressed and sources lie well below the correlation (i.e.\ at lower radio luminosities; see \citealp{Fender04} for a summary of these behaviors). We also note that the range of BH masses for the sample of known XRBs is very small, with a mean of $\approx7$~M$_\odot$ and a standard deviation of $\approx3$~M$_\odot$ \citep[e.g.][]{Ozel2010,Farr2011}, much smaller than the range of galaxy SMBH masses. In this analysis we focus on hard and quiescent XRBs some of which are still detectable even in quiescence. 

\subsection{New ACA observations}
\label{ACA}

 Our sample of XRBs is Cygnus X-1, GRS~1915+105, GX~339-4 and A0620-00. Using the Atacama Compact Array (ACA), we have acquired new mm continuum data of these XRBs (Proposal ID: 2023.1.00887; PI: M. Bureau). We observed the four targets at $230$~GHz ($\approx1$~mm) to match the previous continuum observations of the primary sample that defined the mmFP \citep[see][]{Ruffa24}. The ACA was chosen as XRBs do not have extended emission, so the angular resolution (4$''$.448--6$''$.646) of the observations is irrelevant. The data were reduced using the \textsc{Common Astronomy Software Applications} (\textsc{CASA}) pipeline \citep{McMullin07}, adopting a version appropriate for each dataset and a standard calibration strategy. We reduced the data in same way as described in detail in \cite{Davis22} for the WISDOM galaxy sample.

The continuum images were produced by combining the four continuum spectral windows using the \textsc{CASA} task \textsc{tclean} in multi-frequency synthesis mode. The obtained root-mean-square noise levels are in the range $0.42$ -- $0.60$~${\rm mJy~beam^{-1}}$. Two of the four ACA-observed objects (GRS~1915+105 and A0620-00) are undetected in our continuum maps. The measured mm fluxes and associated luminosities (estimated as described in \citealp{Ruffa24}) of all our targets, including upper limits for non-detections, are listed in Table~\ref{tab:properties}.

\subsection{Archival Observations}

To study how flux variability affects the position of our targets on the mmFP, where possible we additionally gathered archival mm continuum data. The mm luminosity of Cygnus X-1, GRS~1915+105 and GX~339-4 was thus also estimated from archival ALMA continuum observations at $230$~GHz for Cygnus X-1 (Proposal ID: 2016.1.00496.S) and GX~339-4 (Proposal ID: 2019.1.01324.T) and at $104$~GHz for GRS~1915+105 (Proposal ID: 2017.1.00051.S), following the same procedure described in Section~2 of \citet{Ruffa24}. In the case of A0620-00 we took the mm flux from \cite{Gallo2019} which was observed at $98$~GHz (Proposal ID: 2016.1.00773.S). This paper also reported a measurement at $230$~GHz but it was a marginal detection. Due to the impossibility of observing XTE~J1118+480 with ACA (see Section~\ref{ACA}), its mm luminosity was estimated from James Clerk Maxwell Telescope (JCMT) continuum observations at $350$~GHz \citep{Fender01}. 

Archival mm fluxes at frequencies other than $230$~GHz were scaled to that band assuming a flat spectral index ($\alpha=0$). This unavoidably introduces uncertainties in our comparisons. For GRS 1915+105 where the archival mm observation was taken at 104GHz if we assume a spectral index range of -0.15 to 0.15 which has been observed in XRBs \citep[e.g.][]{Fender2000} the scaled flux would range from decreasing by 11.1\% to increasing by 12.6\%. Similarly, for XTE J1118+480 where the archival mm observation was taken at 350GHz assuming the same spectral index range the scaled flux would range from increasing by 6.5\% and decreasing 6.1\%. Finally, for A0620-00 where the archival mm observation was taken at 98GHz assuming the same spectral index range the scaled flux could range from increasing by 13.7\% to decreasing by 12.0\%. Given the observation that XRBs have very flat spectral indices in the hard spectral state and the flux varies minimally when assuming a range of typical spectral indices we believe our assumption of a flat spectral index ($\alpha=0$) to be valid. However, to avoid our spectral index assumption biasing our analysis we primarily use the new homogeneous ACA observations presented in this paper to perform the analyses described in Section~\ref{sec:Results}. 

The intrinsic (absorption-corrected) $2$ -- $10$-keV luminosities ($L_{\rm X,2-10}$) of our targets were derived from X-ray data available in the literature. When possible we used the \textit{Swift}\footnote{\url{https://www.swift.ac.uk/}} catalogue, as for the majority of our sample sources these were the observations most contemporaneous to the ACA ones, thus minimising the impact of temporal flux variability. Specifically, \textit{Swift} data of GX~339-4 were taken just three days after the ACA observations. For Cygnus X-1 and GRS~1915+105, \textit{Swift} observations were taken about one and nine months before the ACA ones, respectively. We note that the \textit{Swift} data were taken in the $0.3$ -- $10$~keV X-ray band, so we scaled the derived luminosities to the $2$ -- $10$~keV energy range assuming a spectral index of $-0.8$ (the mean spectral index reported by \citealp{ReevesTurner00}). \textit{Swift} data are not available for XTE~J1118+480, so its $2$ -- $10$~keV flux was retrieved from the X-ray observations presented in \cite{Gultekin2019}, which were taken in 2000, just two months apart from the available JCMT mm data. There was no recent \textit{Swift} data for A0620-00 which the only observation being taken in 2010 which has a low count. We instead use the \textit{Chandra} observation from \cite{Dincer2017} and use it for both our archival and ACA data points. This observation was taken at $3--9$~keV range so as with the \textit{Swift} data we scale it to the $2--10$~keV range using the $\Gamma$ reported in \cite{Dincer2017} for A0620-00 of 2.07. This data was taken in 2013 meaning that A0620-00 is a source where variability can have an significant impact. To quantify this we use the data from Table 1 of \cite{Dincer2017} and compute the fractional variability, $F_{\rm var}$. We find a $F_{\rm var}$ of 0.83 indicating that A0620-00 is significantly variable. The properties of our targets, including the derived mm and X-ray luminosities, are presented in Table~\ref{tab:properties}.

\subsection{Spectral states}\label{spec_states}
We perform a literature search to determine the spectral state of the 5 sources studied here during both the new and archival mm observations and there corresponding X-ray observations. The spectral states of the sources are listed in column 10 of Table \ref{tab:properties}. We summarise our major finding here. It has been reported that Cygnus X-1 transitioned from a hard to soft state in April 2023 \citep[e.g.][]{Steiner2024} with no further reports of spectral changes. For this reason we classify Cygnus X-1 as being in the soft state at the time of our new mm and corresponding X-ray observations. During the archival observations of Cygnus X-1 it was reported by \cite{Meyer-Hofmeister2020} that the source was in the hard state. It has been reported that during the period of our new mm observations and corresponding X-ray observations GRS 1915+105 was in an obscured state \citep[e.g.][]{Rodriguez2025}. During the archival observations of GRS 1915+105 it was in a relatively soft state as reported by \cite{Zhou2025}. It should be noted that GRS 1915+105 is an unusual source that has stayed in a high state for decades. However, it still undergoes spectral state changed including those that resemble the canonical hard state. It was reported that on the 5th May 2024 that GX339-4 was in a hard state \citep{Russell2024} and as no state change was reported in between then and our new mm and X-ray observations. We then conclude that when our ACA mm observation and corresponding X-ray data were taken GX339-4 was still in the hard state. It was reported by \cite{Remillard2017} that GX339-4 had started a new outburst and observations taken on October 1st 2017 are consistent with it being in the hard state during which our archival observations took place. We then conclude that for our archival mm and X-ray observations GX339-4 was in the hard state. It was reported by \cite{Remillard2000} on the 29th March 2000 that the spectra of XTEJ1118+480 is similar into shape to Cygnus X-1 in its hard state. We then conclude that XTEJ1118+480 was in the hard state during the archival mm and X-ray observations. Finally, A0620-00 has consistently been in  a quiescent state with the last outburst being in 1975 \citep{Elvis1975}. We conclude that for both the new mm and archival mm observations of A0620-00 and the corresponding X-ray observations it was still in the quiescent state.
\begin{table*}
    \centering
    \scriptsize
    \caption{Stellar mass black hole properties}
    \begin{tabular}{lccccccccc}
         \hline
         Object & Distance & $\log_{10}\left(\frac{M_{\rm BH}}{{\rm M}_\odot}\right)$ & $f_{\rm \nu,mm}$ & $\log_{10}\left(\frac{L_{\rm \nu,mm}}{\rm erg~s^{-1}}\right)$ & $\nu_{\rm mm,obs}$ & mm obs.\ date & $\log_{10}\left(\frac{L_{\rm X,2-10}}{\rm erg~s^{-1}}\right)$ & X-ray obs.\ date & State\\
         & (kpc) & & (mJy) & & (GHz) &  & & &\\
         (1) & (2) & (3) & (4) & (5) & (6) & (7) & (8) & (9) & (10)\\
         \hline
         Cygnus X-1 (ACA) & $\phantom{1}1.86$ & $1.17$ & $11.9\pm0.39$ &
         $31.1\pm0.04$ & $230$ & 11-05-2024 & $37.0\pm0.08$ & 08-04-2024 & Soft\\
         
         Cygnus X-1 (Archival) & & & $4.16\pm0.42$ & $30.6\pm0.04$ & $230$ & 06-11-2016 & $36.7\pm0.08$ & 23-11-2016 & Hard\\ 
         
         GRS~1915+105 (ACA) & $11\phantom{.11}$ & 1.00 & $<1.45$ & $<31.7$ & $230$ & 26-05-2024 & $38.1\pm0.08$ & 16-08-2023 & Obscured\\   
         
         GRS~1915+105 (Archival) & & & $2.86\pm0.29$ & $31.6\pm0.04$ & $104$ & 13-10-2017 & $39.7\pm0.08$ & 18-10-2017 & Soft\\
         
         GX~339-4 (ACA) & $10\phantom{.11}$ & $0.76$ & $\phantom{1}7.2\pm0.50$ & $32.3\pm0.04$ & $230$ & 13-06-2024 & $37.1\pm0.08$ & 16-06-2024 & Hard\\
         
         GX~339-4 (Archival) & & & $0.30\pm0.03$ & $30.9\pm0.04$ & $230$ & 13-10-2017 & $37.3\pm0.08$ & 17-10-2017 & Hard\\
         
         XTE~J1118+480 (Archival) & $\phantom{1}1.8\phantom{1}$ & $0.85$ & $4.10\pm0.41$ & $30.7\pm0.04$ & $350$ & 30-05-2000 & $35.5\pm0.08$ & 12-03-2000 & Hard\\
         
         A0620-00 (ACA) & $10\phantom{.11}$ & $0.82$ & $<1.80$ & $<29.7$ & $230$ & 05-12-2023 & $30.7\pm0.08$ & 12-09-2013 & Quiescent\\
         A0620-00 (Archival) & & & $0.044\pm0.008$ & $27.8\pm0.04$ & 98 & 12-11-2016 & $30.7\pm0.08$ & 12-09-2013 & Quiescent\\
         \hline
    \end{tabular}
    \small{\parbox[t]{\textwidth}{\textit{Notes:} (1) Target name, with origin of the data used to derive the quantities in the same row indicated in parentheses. (2) Distance. (3) Black hole mass. (4) -- (7) mm flux, luminosity, frequency and observation date. (8) -- (9) $2$ -- $10$-keV X-ray luminosity and observation date. The distance and black hole mass of Cygnus X-1, GRS~1915+105, XTE~J1118+480 and A0620-00 were taken from \protect\cite{Gultekin2019}, those of GX~339-4 from \protect\citet{Tremou20}. (10) lists the perceived states at the time of observation.}
    } 
    \label{tab:properties}
\end{table*}

\section{Results}
\label{sec:Results}

\begin{figure*}
    \centering
    \includegraphics[width=\textwidth]{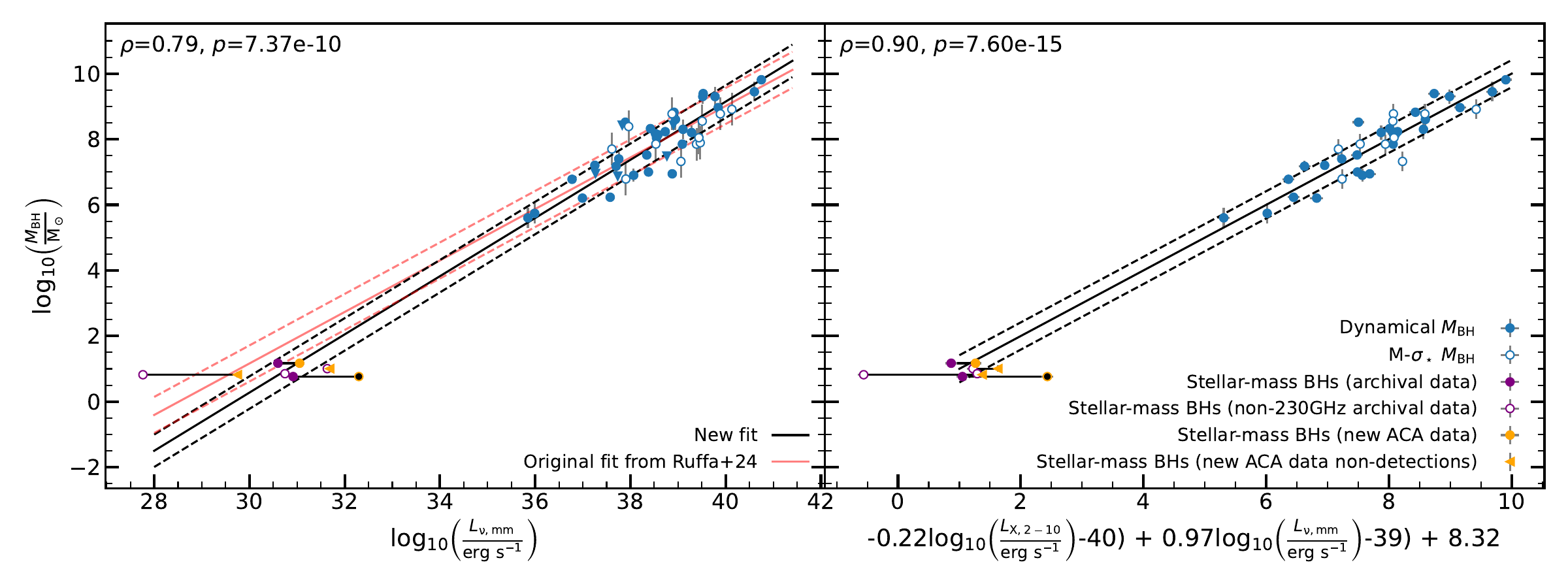}
    \caption[]{Correlation between $M_{\rm BH}$ and $L_{\rm \nu,mm}$ (left panel) and edge-on view of the $M_{\rm BH}$ -- $L_{\rm X,2-10}$ -- $L_{\rm \nu,mm}$ correlation (right panel), as Figure~1 of \citet{Ruffa24}. The four stellar-mass black holes with archival data are plotted as red circles where the objects without 230GHz archival observations are shown as open circles. Green circles (or left-pointing triangles for upper limits) represents $L_{\rm \nu,mm}$ measured using our newly-acquired ACA data. The black filled circle is the ACA observation of GX 339-4. Where possible, a black line connects the original archival data points to the updated ones. Filled blue circles are for SMBH masses estimated dynamically, open blue circles for those estimated using the $M_{\rm BH}$ -- $\sigma_\star$ relation of \cite{Vandenbosch16}. Error bars are plotted for all data points but some are smaller than the symbol used. The best-fitting power-law relations are overlaid as black solid lines, the observed scatters around those relations as black dashed lines. The red solid and dashed lines in the left panel show the best-fitting $M_{\rm BH}$ -- $L_{\rm \nu,mm}$ power-law relation and scatter of \citet{Ruffa24} when including only SMBHs. The correlation coefficients $\rho$ and $p$-values of the Spearman rank analyses are reported in the top-left corner of each panel.
    }
    \label{fig:FP_SM_Fit}
\end{figure*}

Figure~\ref{fig:FP_SM_Fit} shows the four XRBs with ACA data on the $M_{\rm BH}$ -- $L_{\rm \nu,mm}$ plane and the mmFP (green circles for mm detections, left-pointing triangles for mm non-detections). We note that GX~339-4 has greatly increased its mm luminosity in the past $7$ years, moving well beyond the observed scatter of both correlations. The data point corresponding to this source is shown as an black filled green circle in Figure~\ref{fig:FP_SM_Fit}, and this behaviour is discussed further in Section~\ref{sec:discuss}. In the same Figure, we also show the positions of all our targets when calculating their mm luminosity using the available archival data (red circles). The two sources with archival observations at frequencies others than $230$~GHz (GRS~1915+105 and XTE~J1118+480; see Table~\ref{tab:properties}) are plotted as unfilled red circles.

Following \citet{Ruffa24}, we fitted the $M_{\rm BH}$ -- $L_{\rm \nu,mm}$ and mmFP data, including the XRBs, using the \textsc{lts\_linefit} and \textsc{lts\_planefit} routines of \citet{Cappellari2013}, respectively. When performing these fits we only included the new ACA data for Cygnus X-1 and the archival data for GX 339-4 (due to its substantial flux variability in the ACA observations). We do this so we are not fitting non-detections in the ACA observations or sources without 230GHz archival observations where our choice of spectral index used to scale the available flux may bias our results. In the following analyses, we only consider data points with mm luminosities derived from the ACA data, with the exception of GX~339-4 (due to its substantial flux variability). We do not include the archival data for XTE J1118+480 as it did not have archival 230GHz observations and we wanted to ensure our conclusions were not biased by our choice of spectral index used to scale the available flux. As clearly illustrated by Figure~\ref{fig:FP_SM_Fit}, both SMBHs and XRBs follow the same $M_{\rm BH}$ -- $L_{\rm \nu,mm}$ relation and lie on the same mmFP. The resulting best-fitting $M_{\rm BH}$ -- $L_{\rm \nu,mm}$ power law is
\begin{equation}
    \log_{10}\left(\frac{M_{\rm BH}}{\rm M_{\odot}}\right)=(0.89\pm0.04)\left[\log_{10}\left(\frac{L_{\rm \nu,mm}}{\mathrm{erg\,s}^{-1}}\right)-39\right] + (8.26\pm0.08)\,,
\end{equation}
with an observed scatter ($\sigma_{\rm obs}$) of $0.49$~dex and an estimated intrinsic scatter ($\sigma_{\rm int}$) of $0.44\pm0.15$~dex. The best-fitting plane in the ($\log M_{\rm BH}$, $\log L_{\rm X,2-10}$, $\log L_{\nu,\rm mm}$) space is
\begin{eqnarray}
    \nonumber \log_{10}\left(\frac{M_{\rm BH}}{\rm M_{\odot}}\right) =  (-0.22\pm0.05)\left[\log_{10}\left(\frac{L_{\rm X,2-10}}{\mathrm{erg\,s}^{-1}}\right)-40\right] \\+   (0.97\pm0.05)\left[\log_{10}\left(\frac{L_{\nu,\rm mm}}{\mathrm{erg\,s}^{-1}}\right)-39\right] + (8.32\pm0.08)\,,
\end{eqnarray}
with $\sigma_{\rm obs}=0.42$~dex and $\sigma_{\rm int}=0.36\pm0.07$~dex.

In both cases, the best-fitting coefficients agree well with those derived when fitting only the SMBHs \citep{Ruffa24}, and both relations have similar or improved scatters compared to the SMBH-only fits. We also performed Spearman rank analyses to quantify the statistical significance of the correlations, and show the resulting correlation coefficients $\rho$ and $p$-values in the top-left corner of each panel of Figure~\ref{fig:FP_SM_Fit} which are similar or better then the correlation coefficients in \cite{Ruffa24}. The resulting coefficients $\rho$ and $p$-values both suggest a tight and significant correlation between BH mass, mm luminosity and X-ray luminosity.

\section{Physical Drivers}
\label{sec:models}

As illustrated in Figure~\ref{fig:FP_SM_Fit} and discussed above the four XRBs with ACA data (apart from GX 339-4) and the five XRBs with archival data (apart from A0620-00) analysed in this work are in good agreement with the $M_{\rm BH}$ -- $L_{\rm \nu,mm}$ correlation and the mmFP, extending both by $\approx5$ orders of magnitude. This suggests that the dominant mechanism giving rise to their mm continuum emission (and its correlation with their $2$ -- $10$-keV emission) is similar to that in the AGN that were used to discover and define the original correlations (mostly but not exclusively LLAGN). To test this hypothesis, we adopt the same approach as \citet{Ruffa24} and compare the observed mm and X-ray luminosities of the three ACA-observed persistent transients and the archival observations of GX 339-4 to those extracted from mock spectral energy distributions (SEDs). These were constructed using either radiatively-inefficient (ADAF-like) accretion flow or compact radio jet models. As there is currently no evidence for the presence of a dusty torus in XRBs, we do not consider here the associated emission mechanisms (see \citealp{Ruffa24} for the application of such models to SMBHs). For Cygnus X-1 in the soft state and potentially in the hard state there might be some contribution to the mm emission from the donor star wind. This contribution should be investigated further. 

\subsection{ADAF model}
\label{sec:adaf_models}

Advection-dominated accretion flows (ADAFs) are typically described as geometrically-thick two-temperature structures in which the rates of matter accretion onto the black holes are well below the Eddington limits (i.e.\ $\ll\,0.01$~$\dot{M}_{\rm Edd}$; \citealp{NarayanYi1995,Ho08}). \citet{Ruffa24} demonstrated that ADAF-like emission can plausibly explain the existence of the correlations in Figure~\ref{fig:FP_SM_Fit} at high SMBH masses and for all the types of AGN included in their analysis (both high- and low-luminosity sources). To test whether this also applies to stellar-mass BHs, we build model SEDs using the LLAGN model of \citet{Pesce21}, itself a development of previous models by \citet{NarayanYi1995} and \citet{Mahadevan1997}. For full details of how this is implemented and tested, see Section~4.2 of \cite{Ruffa24}. In short, we generate a set of model SEDs for a grid of SMBH masses ($10^{0}$ -- $10^{10}$~M$_{\odot}$) and Eddington ratios $\dot{M}/\dot{M}_{\rm Edd}$ ($10^{-7}$ -- $10^{-2}$), while all the other free parameters are kept at their defaults. We then extract the predicted $230$~GHz and $2$ -- $10$-keV luminosities, and compare them with the measured ones. 

\begin{figure}
    \centering
    \includegraphics[width=\linewidth]{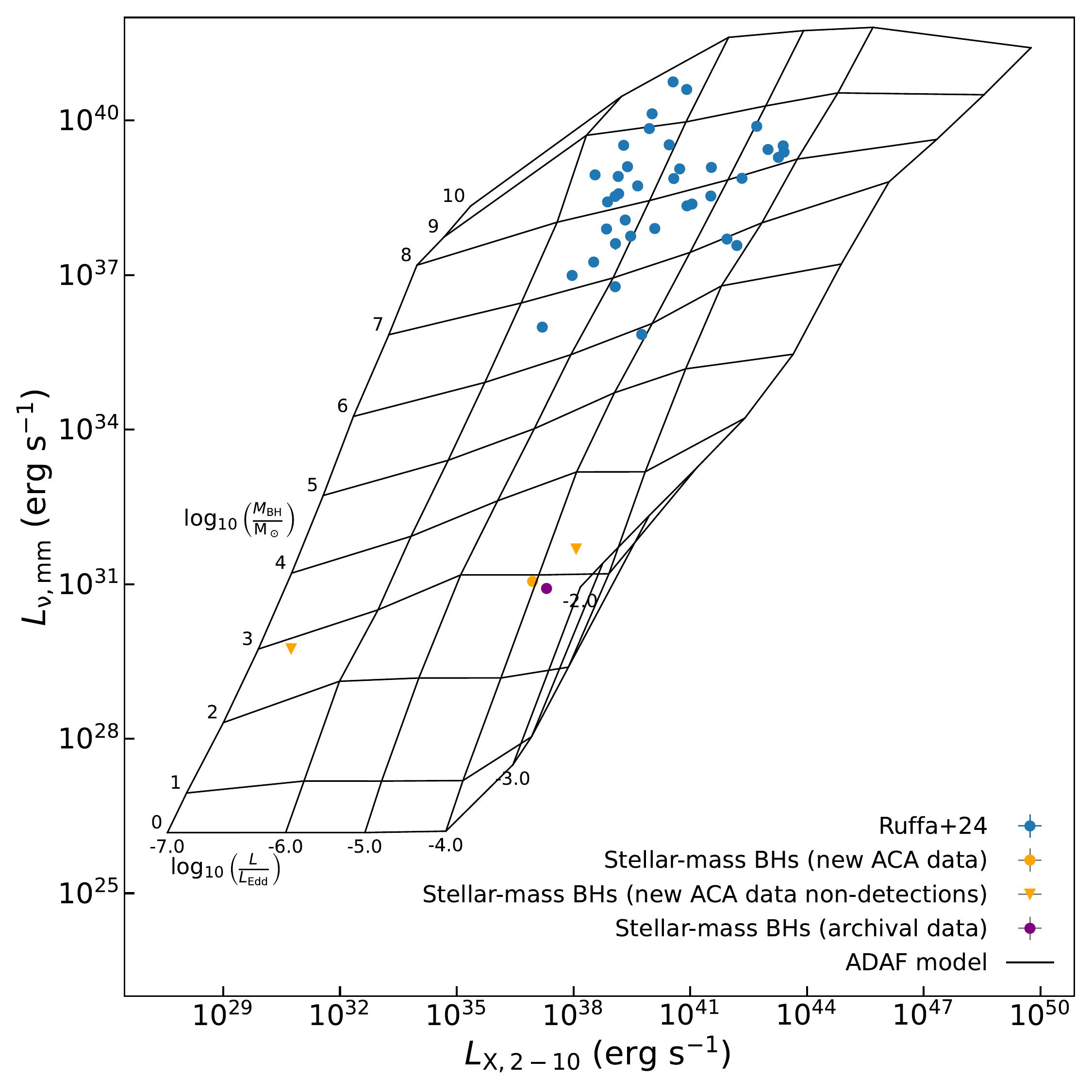}
    \caption{Correlation between $L_{\rm \nu,mm}$ and $L_{\rm X,2-10}$ for both SMBHs (blue data points) and stellar-mass BHs (green and red data points). The black grid illustrates the area covered by the ADAF model solutions as a function of $M_{\rm BH}$ and Eddington ratio $L/L_{\rm Edd}$ (see Section~\ref{sec:adaf_models} for details). We only include data points with mm luminosities derived from the ACA data, with the exception of GX~339-4 (due to its substantial flux variability) where we use the archival data. We not include the archival data for XTE J1118+480 as it does not have archival 230GHz observations.
    }
    \label{fig:FP_SM_grid}
\end{figure}

As shown in Figure~\ref{fig:FP_SM_grid}, all the sources lie within the area covered by the ADAF model solutions (illustrated by the black grid), although these overpredict the BH masses of the persistent transients (listed in Table~\ref{tab:properties}) by about $2$ orders of magnitude. ADAF-like accretion processes have long been believed to occur in hard-state XRBs and to significantly contribute to their emission (see e.g.\ \citealt{Esin97}, \citealt{Narayan98} and Section~\ref{sec:discuss} for more details). Therefore, one possibility to explain this discrepancy is to ascribe it to the large uncertainties of some of the free parameters of the models, such as those describing the (unknown) plasma physics within radiatively-inefficient flows. As mentioned above, we kept all of these parameters fixed to their default values (presented and discussed in Appendix~A of \citealp{Pesce21}, and calibrated only for SMBHs) for our analyses. Our results thus clearly highlight the need to put more robust constraints on all the free parameters of ADAF-like emission models, over the full BH mass range. 

\begin{figure}
    \centering
    \includegraphics[width=\linewidth]{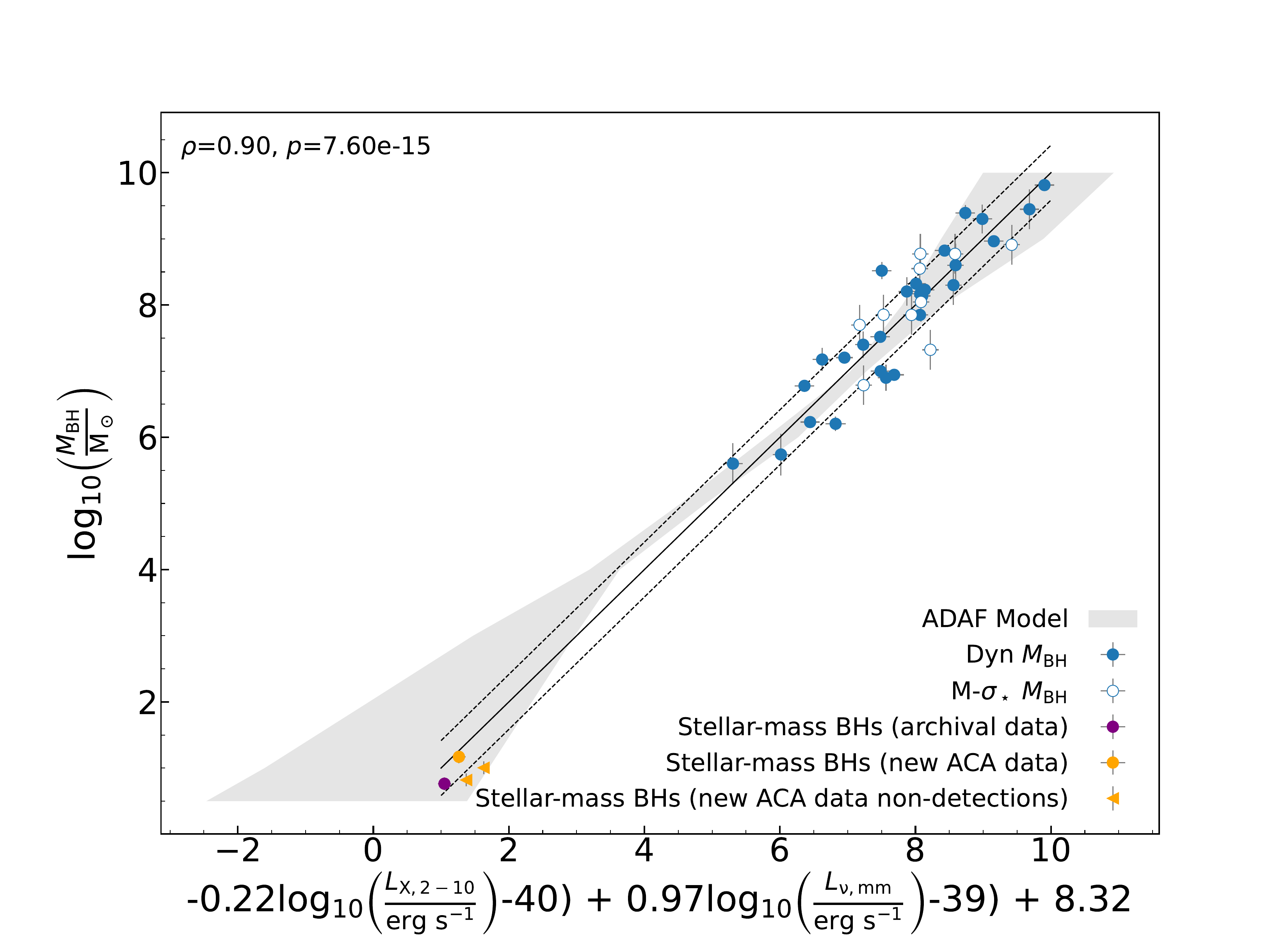}
    \caption{Same as the right panel of Figure \ref{fig:FP_SM_Fit}, but with the grey shaded region representing the projection of the ADAF model grid onto the mmFP. We use the same sample as in Figure \ref{fig:FP_SM_grid}.
    }
    \label{fig:FP_SM_ADAF}
\end{figure}

A second hypothesis to explain the mass inconsistency is that most of the observed emission arises from a different physical process. In Figure~\ref{fig:FP_SM_ADAF}, however, we show the projection of the ADAF model grid onto the best-fitting mmFP, illustrating how these models well reproduce the gradient of the whole correlation. Therefore, while the issues described above prevent us from drawing strong conclusions, we cannot completely exclude that ADAF-like accretion is among the physical mechanisms underlying the mmFP across the entire BH mass regime. If confirmed, our results can also be used to help tune the unknown parameters of the associated emission models.

\subsection{Compact jet model}
\label{sec:compact_jet}

It is currently well established that accreting XRBs in the hard state are able to launch powerful outflows of relativistic particles (i.e.\ jets), which are considered to be scaled-down versions of those identified in the cores of some AGN \citep[see e.g.][]{Fender04b}. The sizes of compact hard state jets in XRBs are usually limited to $\approx 10^{15}$~cm and are often unresolved by radio/mm observations. They are thus typically referred to as compact jets \citep[e.g.][]{Fender04}. Among the few cases in which such compact structures have been spatially-resolved are Cygnus X-1 \citep{Stirling01} and GRS~1915+105 \citep[e.g.][]{Fuchs03}, both part of our sample (see Table~\ref{tab:properties}).

\begin{figure}
    \centering
    \includegraphics[width=\linewidth]{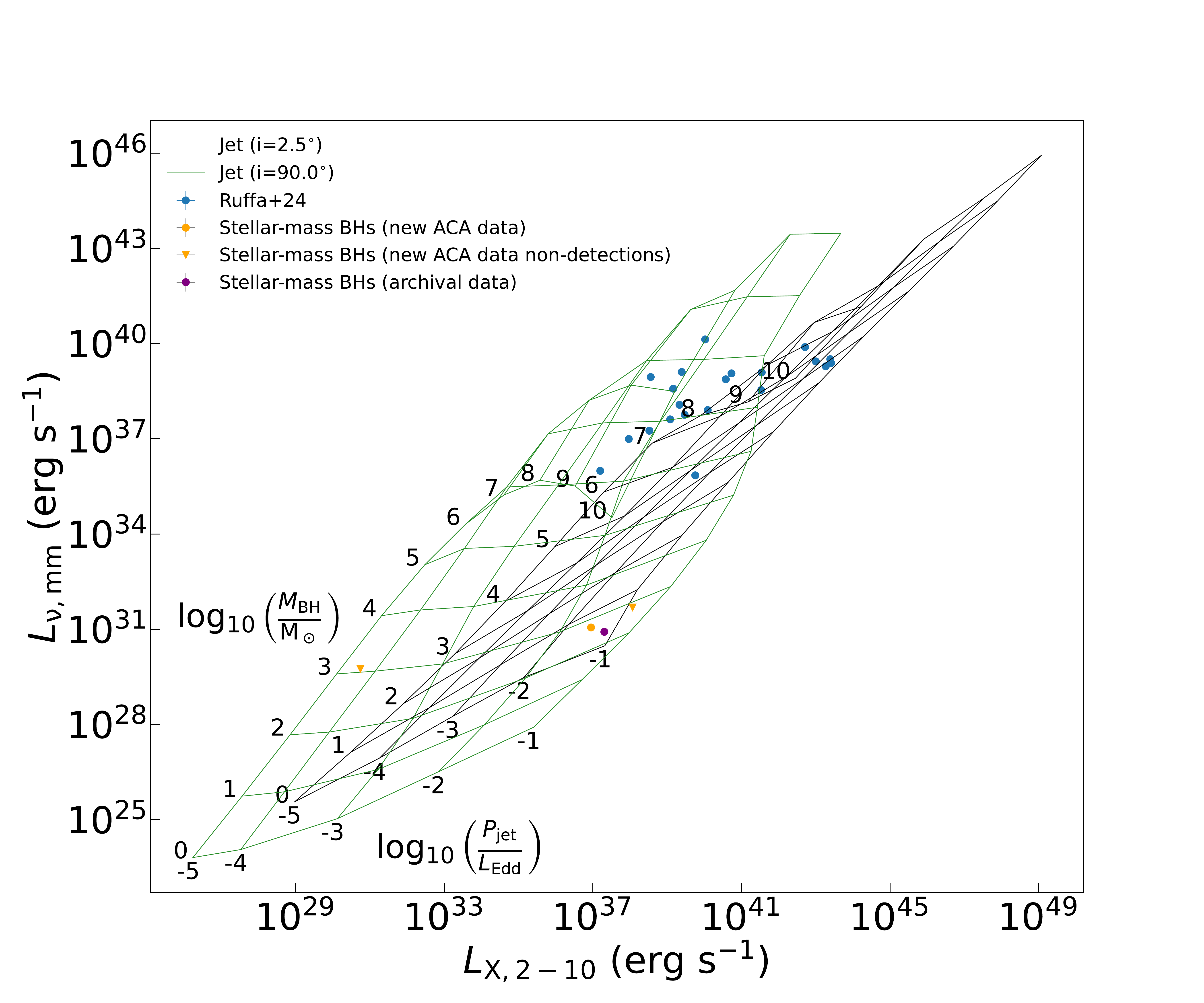}
    \caption{Correlation between $L_{\rm \nu,mm}$ and $L_{\rm X,2-10}$ for both SMBHs (blue data points) and stellar-mass BHs (green and red data points). The black and green grids illustrate the areas covered by the compact jet model solutions as a function of $M_{\rm BH}$ ($10^{1}-10^{10}\,{\rm M_\odot}$) and jet power ($10^{-5.5}-10^{-1.5}{\rm L_{Edd}}$) for jet inclinations of $2.5^{\circ}$ and $90^{\circ}$, respectively. Solutions with intermediate inclinations lie between these two extremes (see Section~\ref{sec:compact_jet} for details). We use the same sample as in Figure \ref{fig:FP_SM_grid}.
    }
    \label{fig:FP_SM_Jet}
\end{figure}

To determine if emission from compact jets can explain the observed trends, we use the {\tt BHJET} model of \cite{Lucchini2022}, extrapolating it down to the stellar-mass BH regime. For full details of how this model is implemented, see Section~4.3 of \cite{Ruffa24}. In Figure\,\ref{fig:FP_SM_Jet}, we show the $L_{\rm X,2-10}$ -- $L_{\rm \nu,mm}$ relation with the resulting jet model grids overlaid in black and green as a function of $M_{\rm BH}$ ($10^{0}-10^{10}\,{\rm M_\odot}$) and jet power ($10^{-5.5}-10^{-1.5}{\rm L_{Edd}}$) for the extremes jet inclinations $2.5^{\circ}$ and $90^{\circ}$, respectively. Jets at intermediate inclinations $i$ lie between these two extremes (but evolve quickly towards the $i=90^{\circ}$ solution once the line-of-sight is no longer aligned with the jet cone). The model solutions clearly encompass all the sources included in our analysis. However, as noted by \citet{Ruffa24} for the SMBH mass regime, the model solutions have significant curvature in the 3D $M_{\rm BH}$ -- $L_{\rm X,2-10}$ -- $L_{\rm \nu,mm}$ space. Therefore, the correlations present in Figure~\ref{fig:FP_SM_Fit} do not seem to occur naturally within the {\tt BHJET} jet model (as projections of the higher-order surface onto the axes).

Furthermore, we note that the model solutions at $i=90^{\circ}$ (green grid in Figure~\ref{fig:FP_SM_Jet}) overpredict $M_{\rm BH}$ of the persistent transients by about the same order of magnitude as the ADAF solutions (see Section~\ref{sec:adaf_models}). The BH masses predicted by the solutions at $i=2.5^{\circ}$ (black grid in Figure~\ref{fig:FP_SM_Jet}) are instead more accurate. However, it is clearly highly unrealistic to assume that all of the sources included in our sample have their compact jets aligned with the line-of-sight. The jet inclination angles of Cygnus X-1 and GRS 1915+105 are estimated to be $\theta\approx35^{\circ}$ and $\theta\approx53^{\circ}$ respectively \citep{Russell07,Zdziarski05}. Such inconclusive results are perhaps unsurprising, however, as compact jet models are substantially more complex than ADAFs, and similarly to them they have many free parameters (that we again left at their default values) that are essentially unconstrained. Future work will thus be crucial to better constrain the plasma physics in these models and further test our results.

\section{Discussion and conclusions}
\label{sec:discuss}

\cite{Ruffa24} reported the discovery of tight $M_{\rm BH}$ -- $L_{\rm \nu,mm}$ and $M_{\rm BH}$ -- $L_{\rm X,2-10}$ -- $L_{\rm \nu,mm}$ correlations. The latter has been dubbed the "mm fundamental plane of BH accretion" and holds for both high- and low-luminosity AGN. \cite{Ruffa24} also investigated potential physical mechanisms underlying the nuclear emission of the sample sources and thus driving the newly-discovered correlations. According to the standard paradigm, a high-luminosity (i.e.\ high-accretion rate) AGN should accrete through a classic geometrically-thin and optically-thick accretion disc surrounded by a dusty torus \citep[e.g.][]{Heckman14}. In this scenario, both the mm and the $2$ -- $10$-keV emission arise from the accretion disc, reprocessed by torus dust in the mm and Compton up-scattered by the hot corona in the X-rays. In a typical low-luminosity (i.e.\ low-accretion rate) AGN, a classic accretion disc should be either absent or truncated at some inner radius (the transition usually happening beyond a few tens of Schwarzschild radii), and should be replaced by some sort of ADAFs \citep{Nara95,Ho08}. \citet{Ruffa24} discovered instead that, for both high- and low-luminosity sources, the observed correlations are best explained if the nuclear emission in the mm and X-rays primarily arises from an ADAF-like process, as the torus models underpredict the nuclear mm emission ($L_{\rm \nu,mm}$) of high-luminosity AGN by at least $2$ orders of magnitude. This provides support to some accretion disc models, allowing discs to transition from ADAF-like to geometrically thin (and vice versa at different radii), and for ADAFs to exist above and below classic accretion discs \citep{Mahadevan1997}.

\citet{Ruffa24} also explored the possibility that both the mm and the X-ray emission arises from compact (and thus probably young; \citealp{ODea20}) radio jets. These jets have been argued to dominate the whole SEDs of LLAGN \citep[e.g.][]{Fernandez-Ontiveros2023} and have spectral properties similar to those of ADAFs at radio and mm wavelengths. These types of models are however only marginally consistent with the observed correlations \citep{Ruffa24}.  

In this work, we extended both the $M_{\rm BH}$ -- $L_{\rm \nu,mm}$ relation and the mmFP to the stellar-mass BH regime. The main aim was to establish whether these correlations also hold for accreting stellar-mass BHs, and if so whether ADAF models are able to reproduce the observed trends for all types of sources. For this, we used available archival and new dedicated ACA observations at $230$~GHz of all known transient XRBs that are still detectable in quiescence (i.e.\ LLAGN-equivalent) ($5$ objects in total; see Table~\ref{tab:properties}). As illustrated in Figure~\ref{fig:FP_SM_Fit}, the five XRBs included in our analysis tightly follow both correlations, as they do at radio wavelengths \citep[e.g.][]{Merloni03,Falcke04,Plotkin12}. Two sources lie significantly outside the relation. Firstly, GX~339-4 is significantly outside of the observed scatter of both correlations when its $L_{\rm \nu,mm}$ is inferred using our newly-acquired ACA data, whereas it is well within the scatter when $L_{\rm \nu,mm}$ is estimated using archival data (see Table~\ref{tab:properties}). This is likely because the mm luminosity of this source increased greatly since the previous mm observations in 2017 (see Table~\ref{tab:properties}). This could be due to this source shifting to a soft accretion state and therefore now deviates from the correlations, as a similar behavior exists in the radio FP when XRBs change state \citep[see e.g.][]{Merloni03,Gultekin09,Gultekin2019},  however, this is not supported by the published states of GX~339-4 where in both the ACA and archival observation it was in the hard state (see Section \ref{spec_states} and Table \ref{tab:properties}). Secondly, the archival data of A0620-00 lies significantly out side of the scatter of both relations due to its very low mm luminosity. Its corresponding ACA upper limit, however, lies in line with both relations. A0620-00 is a unique source that has been in quiescence since its last outburst in 1975 \citep{Elvis1975} so we speculate it could be a "unique" state which could explain why it does not follow these relations like other XRBs appear to do.  We also note that the lack of simultaneous X-ray observations for some sources may further increase the observed scatter. 

Our results suggest that the dominant mechanism giving rise to the mm continuum emission of hard-state XRBs (and its correlation with the $2$ -- $10$-keV emission) is similar to that in the AGN originally used to define the correlations. To test this hypothesis, we adopted the same approach of \citet{Ruffa24} and extended the ADAF and compact jet models of \citet{Pesce21} and \citet{Lucchini2022}, respectively, down to the stellar-mass BH regime.

In radiatively-inefficient flows such as ADAFs, the electrons cool down via a combination of self-absorbed synchrotron, Bremsstrahlung and inverse Compton radiation, which together yield an ADAF spectrum near the BHs \citep[e.g.][]{Narayan98}. These types of accretion solutions were first described by \citet{Ichimaru77}, who also first suggested that the observed spectra of hard-state XRBs such as Cygnus X-1 are likely dominated by emission from ADAF-like accretion processes. This hypothesis was later supported by a number of works (see e.g.\ \citealt{Esin97}, \citealt{NarayanYi1995} and references therein), which also demonstrated the applicability of ADAFs to other types of systems, especially typical LLAGN \citep{Narayan98}. Compact jets produce self-absorbed synchrotron emission which manifests itself as a flat (spectral index $\alpha\approx0$)
or inverted ($\alpha\gtsimeq0$) spectral component at radio and (sub-)millimetre bands \citep[e.g.][]{Fender01,Fender04,Done07,Fender09,Kylafis12,Sikora23}. The resulting spectrum is therefore very similar to that expected from an ADAF \citep[e.g.][]{Narayan98}. It has also been demonstrated that the radio luminosities of compact jets from XRBs show a strong, non-linear correlation with their X-ray luminosities (e.g.\ \citealp{Corbel03} and \citealp{Gallo03} and references therein). Both the ADAF and compact jet models can potentially give rise to the observed emission down to stellar BH masses, although they mostly overpredict the XRB masses by about $2$ orders of magnitude. We speculate that this is caused by the large uncertainties of the parameters describing the plasma physics in both models, most of which have so far been calibrated only for SMBHs. This clearly prevents us from drawing strong conclusions on this issue at this stage. We note that these conclusions differ from those of \cite{Ricci23} who found that the driving mechanism behind the mm emission is likely due to the X-ray corona. This was discussed in \cite{Ruffa24} where the mmFP was introduced.

Future modeling work is needed to test whether a hybrid jet-ADAF model can also predict the emission we are seeing and the reproduce the mmFP. In these models the X-rays could be emitted by the ADAF and the mm emission from the jet. Further observational work needs to be carried out to fully constrain where stellar-mass black holes are on the newly-discovered correlations, and how variability affects this. Further theoretical work is also critical to fully understand the mechanism driving these correlations. Such work will be invaluable, shedding light on the physics of accretion across mass scales, and revealing the common processes that apply both within our galaxy and in extragalactic systems.

\section*{Acknowledgements}

JSE is supported by the UKRI AIMLAC CDT, funded by grant EP/S023992/1. IR and TAD acknowledge support from grant ST/S00033X/1 through the UK Science and Technology Facilities Council (STFC). MB, TGW and RF were supported by STFC consolidated grant `Astrophysics at Oxford' ST/H002456/1 and ST/K00106X/1. JG gratefully acknowledges funding via STFC grant ST/Y001133/1.   
This paper makes use of ALMA data. ALMA is a partnership of ESO (representing its member states), NSF (USA) and NINS (Japan), together with NRC (Canada), NSC and ASIAA (Taiwan), and KASI (Republic of Korea), in cooperation with the Republic of Chile. The Joint ALMA Observatory is operated by ESO, AUI/NRAO and NAOJ.
This paper also made use of the NASA/IPAC Extragalactic Database (NED) which is operated by the Jet Propulsion Laboratory, California Institute of Technology under contract with NASA. We also acknowledge usage of the HyperLeda database. This paper made up work towards the PhD thesis \cite{JacobElfordThesis}.

\section*{Data Availability}

The raw ALMA data used in this article are all available to download at the ALMA archive (\url{https://almascience.nrao.edu/asax/}). The calibrated data, final products and original plots generated for this research study will be shared upon reasonable request to the first author. The X-ray data were retrieved from the catalogue of \citet{Bi20} or from the NASA/IPAC Extragalactic Database (NED; \url{https://ned.ipac.caltech.edu/}).



\bibliographystyle{mnras}
\bibliography{mybibliography} 




\appendix


\bsp	
\label{lastpage}
\end{document}